\documentclass{myws-p8}

\usepackage{amsmath,graphicx}

\newcommand{\mN}{m_{\scriptscriptstyle N}}
\newcommand{\mD}{m_{\scriptscriptstyle \Delta}}

\begin{document}

\title{Baryon Chiral Dynamics\footnote{\MakeLowercase{\MakeUppercase{T}alk given at the 9th \MakeUppercase International \MakeUppercase Conference on the \MakeUppercase Structure of \MakeUppercase Baryons (\MakeUppercase Baryons 2002), \MakeUppercase Newport \MakeUppercase News, \MakeUppercase Virginia, \MakeUppercase March 3-8, 2002.}}}

\author{Thomas Becher}

\address{Stanford Linear Accelerator Center\\
Stanford University, Stanford, CA 94309, USA\\	
E-mail: tgbecher@SLAC.stanford.edu}

\maketitle

\vspace*{-5cm}
\begin{flushright}
SLAC-PUB-9268 
\end{flushright}
\vspace*{5cm}

\abstracts{After contrasting the low energy effective theory for the
baryon sector with one for the Goldstone sector, I use the example of pion nucleon scattering to discuss some of the progress and open issues in baryon chiral perturbation theory.}

\section{Higher, faster, swifter?}

Many of the constraints of chiral symmetry on the interaction of pions
and nucleons at low energies were worked out long before the advent of
QCD (Current algebra, PCAC). Later it was realized that the
corrections to these symmetry relations can be obtained by
implementing the chiral symmetry and its breaking by the quark masses
into an effective Lagrangian describing the interaction of mesons and
baryons. This method is called chiral perturbation theory
(CHPT).\cite{gl} It allows one to compute the expansion of QCD
amplitudes and transition currents in powers of the external momenta
and quark masses; it has become one of the standard tools to analyze
the strong interactions at low energy.

Over the last few years, the progress in this field in the baryon
sector has been twofold: on one hand, we have reached a new level of
precision in many of the classical applications: by now, the full one
loop result for the nucleon form factors and the pion-nucleon
scattering amplitude in the isospin limit is
known.\cite{Fettes:2000xg,Becher:2001hv} Even the first two-loop
result has been obtained: the chiral expansion of the nucleon mass has
been worked out to fifth order.\cite{McGovern:1998tm} On the other
hand, the framework has been extended and applied to a whole range of
new processes: the effective Lagrangian has been extended to include
electromagnetism, making it possible to disentangle strong and
electromagnetic isospin violation.\cite{Muller:1999ww} This effective
theory of QCD+QED has been used to calculate next-to-leading order
isospin violating effects in the pion-nucleon scattering
amplitude\cite{isospin} and to study the properties of the $\pi^- p$ bound
state.\cite{Gasser:2002am} Another extension of the framework
incorporates the $\Delta$-resonance as an explicit degree of freedom
into the effective Lagrangian, thereby summing up the potentially
large higher-order terms in the chiral expansion associated with this
resonance.\cite{Hemmert:1997ye}

Despite all of this impressive progress, we are still short of having
the answers to some very old questions, like, for example, what is the
value of the $\sigma$-term, and the agreement with data in many cases
is not quite as satisfactory as in the Goldstone sector. In my talk, I
will contrast the baryon with the meson sector and illustrate some of
the peculiarities that arise, once the baryon field is included into
the effective Lagrangian.

I will illustrate my discussion with the example of $\pi N$-scattering
and conclude that the low energy theorems for this amplitude hold to
high accuracy.  Chiral symmetry governs the amplitude in a small
region around the Cheng-Dashen point. However, the momentum dependence
of the chiral representation for the amplitude is not accurate enough
to make direct contact with experimental data. After discussing some
of the difficulties associated with the extrapolation of the
experimental results to the low energy region, I show how the simple
structure of the result in the low energy theory can be implemented
into a dispersive analysis of the data.

As I focus the discussion mostly on $\pi N$-scattering, I will fail to
report on many important developments over the past few
years. Fortunately, my sense of guilt for omitting electromagnetic
probes of the nucleon was relieved by the plenary talks of Ed Brash,
Helene Fonvieille, Frank Maas and Harald Merkel as well as a
number of interesting talks on these matters in the parallel
sessions. Unfortunately, there were no talks covering the few nucleon
sector\cite{Griesshammer:2001ej}, nor about the recent work on
quenched\cite{Labrenz:1996jy} and partially quenched\cite{Chen:2001yi}
baryon CHPT.

\section{Baryons versus Mesons}

While this is not the place to give an introduction to
CHPT\cite{overviews}, it is instructive to point out some of the
differences between CHPT in the baryon and the Goldstone sector.  All
in all, the inclusion of the baryon field leads to three
complications: i) in general, one has to deal with a larger number of
low energy constants than in the vacuum sector, ii) from the viewpoint
of the low energy theory, the physical region is at higher
energies, iii) the singularity structure of the amplitudes is more
complicated. On the upside, there are much more and more precise data
available than in the meson sector.

\subsection{Effective Lagrangian}
For vanishing up- and down-quark masses, the pions are the Goldstone
bosons associated with the spontaneous breaking of chiral
symmetry. The interactions between Goldstone bosons tend to zero at
low energies and they decouple from matter fields. Accordingly, the
effective Lagrangian is organized in powers of derivatives on the
Golstone fields. At low energies, the terms with higher powers of
derivatives on the meson field are suppressed by powers of the meson
momenta. Because of decoupling, the interactions of the baryon and the
meson involve at least one derivative on the meson field.  The
lowest-order, effective Lagrangian for the pion-nucleon interaction
reads
\begin{equation}
  {\cal L}_{\mbox{\tiny eff}}^{\mbox{\tiny N}}= 
- \frac{g_{A}}{2 F_\pi}\,  
\bar{\psi}\,\gamma^\mu \gamma_5 \,\partial_\mu \pi\psi 
+\frac{1}{8 F_\pi^2}\,\bar{\psi}\,\gamma^\mu i[\pi,\partial_\mu\pi]\psi +\ldots
\end{equation}
The ellipsis stands for terms which involve higher powers of the pion
field. Their coefficients are fixed by chiral symmetry. At
second order, the effective Lagrangian also contains terms
proportional to the quark masses.

The fact that the lowest-order Lagrangian is fully determined by the
nucleon mass and the matrix element of the axial charge shows how
chiral symmetry constrains the interactions of mesons and
baryons. However, the rapid increase in the number of parameters at
higher orders makes it evident that it is not nearly as restrictive as
in the meson sector. The number of parameters entering at each order
is shown in brackets:
\begin{equation*}
\begin{aligned}
{\cal L}_{\pi\pi}&=& &&{\cal L}_{\pi\pi}^{(2)}\phantom{o} &&+&&{\cal L}_{\pi\pi}^{(4)} &&& + &&& {\cal L}_{\pi\pi}^{(6)} \\ 
 &&&& (2)\phantom{o} &&&& (7)\phantom{o} &&&&&& (53) \\ 
{\cal L}_{\pi N}&=&{\cal L}_{\pi N}^{(1)}\phantom{o}&+&{\cal L}_{\pi N}^{(2)}\phantom{o}&+&{\cal L}_{\pi N}^{(3)}\phantom{o}&+&{\cal L}_{\pi N}^{(4)}&&&& \\
	&& (2)\phantom{o} && (7)\phantom{o} && (23)\phantom{o} && (118) &&  &&
\end{aligned}
\end{equation*}
The larger number of low energy constants arises from the
spin-$\frac{1}{2}$ nature of the nucleon and because it stays
massive in the chiral limit, so that the effective Lagrangian involves odd as well as even powers in the chiral expansion. 

In a given process only a handful of the outrageous number of terms in
the fourth-order Lagrangian will contribute. The fact that the
effective Lagrangian contains 118 terms at fourth order,\cite{Fettes:2000gb} however, means that the chances that the same
combination enters two different observables are rather dim: there
will hardly be any symmetry relations valid to fourth
order in the chiral expansion.

\subsection{The low energy region and the role of resonances}

\begin{figure}[thb]
\includegraphics[width=1.0\textwidth]{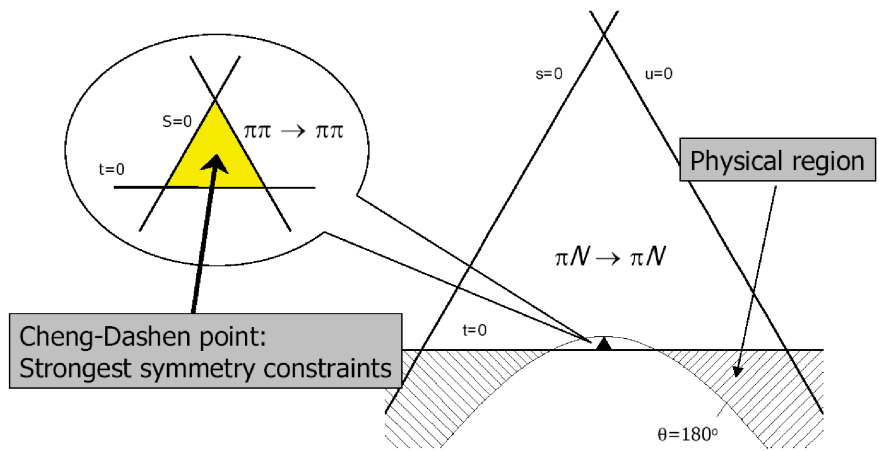}
\caption{Comparison of the Mandelstam triangles for $\pi\pi$- and $\pi N$-scattering.}\label{fig:mandelstam}
\end{figure}

The strongest constraints from chiral symmetry on the $\pi
N$-scattering amplitude are obtained at unphysically small values
of Mandelstam variables, at the Cheng-Dashen point $s=u=\mN^2$,
$t=2\,M_\pi^2$. In figure \ref{fig:mandelstam} the Mandelstam
triangles for $\pi\pi$- and $\pi N$-scattering are compared. The
figure makes it evident that the physical threshold for $\pi
N$-scattering is at higher energies: the increase in $s$ from the
Cheng-Dashen point to the threshold is of $O(M_\pi)$ for $\pi
N$-scattering, while it is $O(M_\pi^2)$ for $\pi\pi$ scattering.
At threshold, higher-order terms in the chiral expansion will
therefore be more important in $\pi N$- than in $\pi \pi$-scattering.

This observation is confirmed by looking at the position of the first
resonance. The increase in $s$ from the Cheng-Dashen point to the
first resonance is roughly the same in both cases:
$\mD^2-m_N^2\approx m_\rho^2$. The relevant expansion parameter
for the resonance contributions at threshold is, however, much larger
for $\pi N$-scattering: $2\mN\,M_\pi/(\mD^2-\mN^2)\approx 0.4 \gg
4M_\pi^2/m_\rho^2 \approx 0.1$. While the effective theory for the
meson sector will still yield meaningful results well above threshold, the
$\Delta$-resonance must be included into the Lagrangian if one wants
to arrive at an accurate description of the meson nucleon amplitude
above threshold.

This can be done in a systematic way by counting the mass difference
$\delta=\mD-\mN$ as a small quantity of the same order as
$M_\pi$. This procedure is referred to as ``small scale expansion''
and allows one to resum the potentially large corrections associated
with the resonance.\cite{Hemmert:1997ye} While it is certainly
important to get a handle on the resonance contributions, a few words
of caution are appropriate: we are still performing a low energy
expansion and there are higher-order terms not associated with the
resonance. In particular, the inclusion of the $\Delta$ has so far
only be performed in the non-relativistic framework for baryon CHPT
(to be discussed in the next subsection) and the higher-order
kinematic corrections are important already in the threshold
region. Furthermore, the effective theory which includes the $\Delta$
is not unique: recently it has been claimed that the effective
Lagrangian is compatible with different counting schemes; it seems in
general not possible to decide from first principles at which order a
given operator enters.\cite{Hemmert:2002uh} The $\pi N$-scattering
amplitude has been calculated to third order in this combined
expansion in $\delta$ and the meson momenta.\cite{Fettes:2000bb} The
calculation confirms that the bulk of the $\Delta$ contribution stems
from the resonance pole term. According to the authors, the energy
range in which their results reproduces the existing data is only
slightly larger than for the fourth-order calculation in pure CHPT.

\subsection{Formulation of the effective theory}

In the low energy expansion, the baryon four momentum $P_\mu$ has to
be counted as a large quantity, since $P^2=\mN^2$ is of the size of
the typical QCD scale squared. If we choose a frame, where the baryon
is initially at rest and let it interact with low energy pions, the
nucleon will remain nearly static, its three momentum being of the
order of the meson mass. The chiral expansion of the corresponding
amplitudes in the momenta and masses of the mesons therefore leads to
an expansion of the nucleon kinematics around the static limit. This
expansion is implemented ab initio in the framework called heavy
baryon chiral perturbation theory (HBCHPT).\cite{hbchpt} However, the
expansion of the kinematics fails to converge in part of the low
energy region. The breakdown is related to the fact that the expansion
of the nucleon propagator in some cases ruins the singularity
structure of the amplitudes. This makes it desirable to perform the
calculations in a relativistic framework. In doing so, the correct
analytic properties of the amplitudes are guaranteed, and one can
address the question of their chiral expansion in a controlled way.

In the relativistic formulation of the effective theory a technical
complication arises from the fact that in a standard regularization
prescription, like dimensional regularization, the low energy
expansion of the loop graphs starts in general at the same order as
the corresponding tree diagrams.\cite{gss} Since the contributions
that upset the organization of the perturbation expansion stem from
the region of large loop momentum of the order of the nucleon mass,
they are free of infrared singularities. In $d$-dimensions, the
infrared singular part of the loop integrals can be unambiguously
separated from the remainder, whose low energy expansion to any finite
order is a polynomial in the momenta and quark masses. Moreover, the
infrared singular and regular parts of the amplitudes separately obey
the Ward identities of chiral symmetry. This ensures that a suitable
renormalization of the effective coupling constants removes the
infrared regular part altogether, so that we may drop the regular part
of the loop integrals and {\em redefine} them as the infrared singular
part of the integrals in dimensional regularization, a procedure
referred to as {\em infrared regularization}.\cite{Becher:1999he} The
representation of the various quantities of interest obtained in this
way combines the virtues of HBCHPT and the relativistic formulation:
both the chiral counting rules and Lorentz invariance are manifest
at every stage of the calculation.

In the meantime, this relativistic framework has been used to calculate the
scalar\cite{Becher:1999he}, axial\cite{julia} and electro-magnetic
form factors\cite{Kubis} as well as the elastic pion-nucleon
amplitude\cite{Becher:2001hv} to fourth order in the chiral
expansion. Recently, the Gerasimov-Drell-Hearn sum rule has been
reanalyzed and it was found that the recoil corrections, which are
summed up in the relativistic approach, are rather
large.\cite{Bernard:2002bs}

\section{Pion-nucleon scattering}

\subsection{Low energy theorems}

Chiral symmetry constrains the strength of the $\pi N$-interaction as
well as the value of the scattering amplitudes at the Cheng-Dashen
point. The fourth-order result for the scattering amplitude allows us
to analyze the corrections to the low energy theorems that arise at
leading order in the expansion and we find that the
symmetry breaking corrections are rather small. 

As a first example, let us consider the Goldberger-Treiman relation
\begin{equation*}
g_{\pi N} = \frac{g_A\,\mN}{2\,F_\pi}(1+\Delta_{\text{GT}})\, .
\end{equation*}
If the masses of the up- and down-quarks are tuned to zero, the
strength of the $\pi N$ interaction is fully determined by $g_A$ and
$F_\pi$: $\Delta_{\text{GT}}=0$. Up to and including terms of third order in $M_\pi$, the correction has the form
\begin{equation*}
\Delta_{\text{GT}}=c\,M_\pi^2+O(M_\pi^4)\, .
\end{equation*}
It is remarkable that the correction neither involves a term of the
form $M_\pi^2\,\ln(\frac{M_\pi^2}{\mN^2})$ (a ``chiral logarithm'')
nor a correction of order $M_\pi^3$. Such infrared singular terms {\em
are} present in the chiral expansion of $g_{\pi N}$, $g_A$, $F_\pi$
and $\mN$, but they cancel out in the above relation. To this order,
the correction is thus analytic in the quark masses. If the low energy
constant $c$ is of typical size, $c\approx 1/\text{GeV}^2$, the correction to
the Goldberger-Treiman relation is $2\%$. If one evaluates the above
relation with value for the coupling constant given
in H\"ohler's comprehensive review of $\pi N$-scattering\cite{hoehler}, one finds $\Delta_{\text{GT}}=4\%$. The data
accumulated since then seems to favor a smaller value of $g_{\pi N}$ reducing the correction to 2-3\%.

Another well known low-energy theorem relates the value of the
isosymmetric amplitude $D^+$ at the Cheng-Dashen point\footnote{The bar indicates that the pseudo-vector Born term has been subtracted.}
\begin{equation*}
\Sigma=F_\pi^2 \bar D^+(s=\mN^2,t=2\,M_\pi^2)
\end{equation*}
to the scalar form factor
\begin{equation*}
\langle N'|\,m_u \,\bar{u}u+m_d\,\bar{d}d\,|N\rangle=
\sigma(t)\,\bar{u}' u\, .\end{equation*}
 The relation may be written in the form
\begin{equation*}
\Sigma= \sigma(2M_\pi^2)+\Delta_{\scriptscriptstyle
CD}\, .
\end{equation*}
The theorem states that the term
$\Delta_{\scriptscriptstyle CD}$ vanishes up to and including
contributions of order $M_\pi^2$.  The explicit expression obtained for
$\Sigma$ when evaluating the scattering amplitude to order $q^4$ again
contains infrared singularities proportional to $M_\pi^3$ and $M_\pi^4\ln
M_\pi^2/\mN^2$. Precisely the same singularities, however, also show up in
the scalar form factor at $t=2M_\pi^2$, so that the result for
$\Delta_{\scriptscriptstyle CD}$ is free of such
singularities: 
\begin{equation*} 
\Delta_{\scriptscriptstyle
CD}=d\, M_\pi^4+O(M_\pi^5) \, .
\end{equation*}
A crude estimate like the one used in the case of the
Goldberger-Treiman relation indicates that the term
$\Delta_{\scriptscriptstyle CD}$ must be very small, of order 1
MeV. Unfortunately, the experimental situation concerning the magnitude
of the amplitude at the Cheng-Dashen point leaves much to be
desired. The inconsistencies between the results of the various
partial wave analyses need to be clarified in order to arrive at a
reliable value for $g_{\pi N}$. Only then it will be possible to
extract a small quantity like the $\Sigma$-term from
data.\footnote{Jugoslav Stahov has reported at the conference that
discrepancies in the higher partial waves of different
partial wave analyses can explain the inconsistencies between different
determinations of the $\Sigma$-term.\cite{Stahov:2002vs}}

\subsection{Momentum dependence: analyticity and unitarity}

To obtain the amplitudes in the region around the Cheng-Dashen point,
the experimental results need to be extrapolated to the subthreshold
region. The extrapolation can only be performed reliably, if the
correct structure of the singularities of the amplitude is implemented
into the data analysis. Having to deal with functions of two
variables, this is not a simple task and while all modern partial wave
analyses incorporate some of these constraints, subsequent analyses
have not kept up with the high level of sophistication reached by the
Karlsruhe-Helsinki collaboration in the eighties.

Because of the complexity of a dispersive analysis, it is tempting to
use the representation obtained in chiral perturbation theory to
perform the extrapolation to the unphysical region, since the use of
a relativistic effective Lagrangian guarantees the correct analytic
properties in the low energy region. The problem with this approach is
that unitarity is not exact in the chiral representation, but only 
fulfilled to the order considered. At one loop level, the imaginary
part will be given by the current algebra amplitudes squared. Since
the corrections to the current algebra result become sizeable above
threshold, the violation of unitarity will prevent an accurate
extrapolation to the subthreshold region in this framework.

\begin{figure}
\begin{center}
\includegraphics[width=0.7\textwidth]{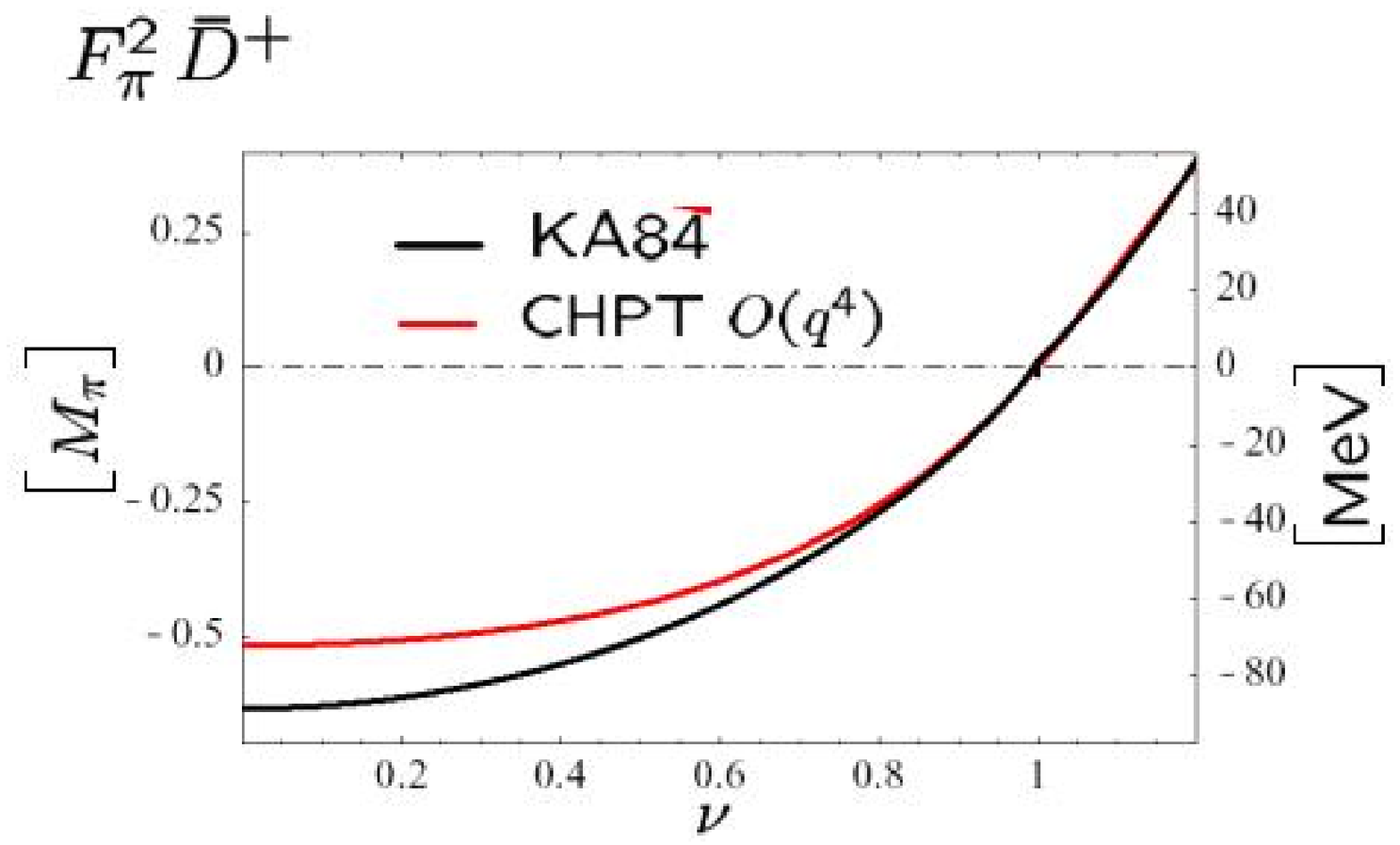} \\
\includegraphics[width=0.64\textwidth]{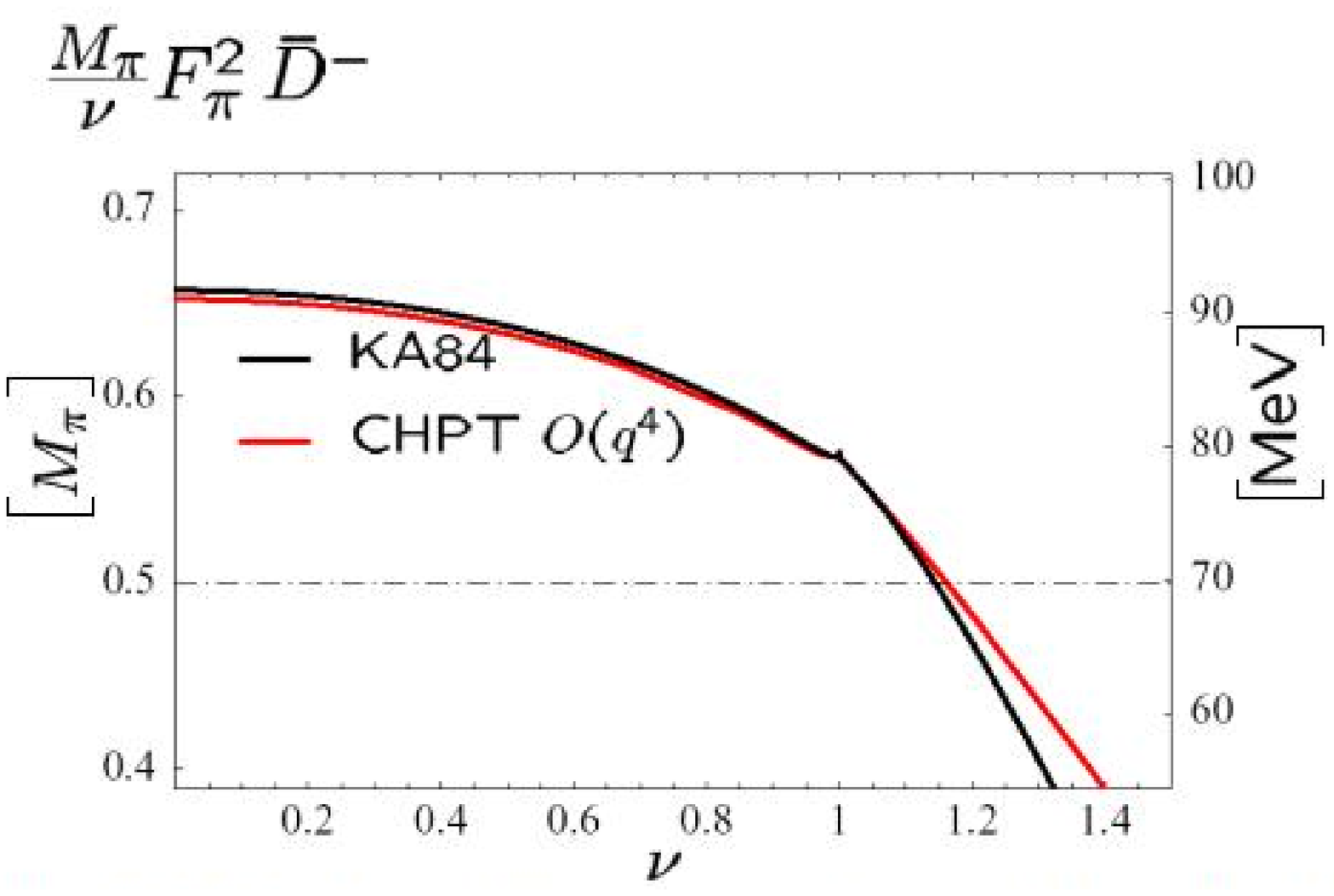}
\end{center}
\caption{Real part of the pion-nucleon amplitude at zero momentum
transfer. The variable $\nu$ denotes the lab.~energy of the incoming
pion. The reaction threshold is at $\nu=M_\pi$. The red line is the
result obtained at the fourth order in the chiral expansion. The black
curve corresponds to the KA84 solution.}\label{fig:momdep}
\end{figure}

This is illustrated in figure \ref{fig:momdep}, where we compare the
result obtained in CHPT with the KA84 solution.\cite{KA84} The
parameters in the chiral representation have been adjusted to the KA84
solution at the threshold and we want to check the energy range in
which we reproduce the KA84 solution. For the amplitude $D^+$, the
deviation in the region around the Cheng-Dashen point would translate
into a $10$ MeV uncertainty in the $\Sigma$-term. The accuracy is
better in the case of the amplitude $D^-$, but also in this case the
chiral representation starts to deviate soon after threshold.

There are various prescriptions\cite{unitarization} to fix the problem
by hand: one can, e.~g.~,use the K-matrix formalism to unitarize the
amplitudes found in CHPT. Once some resonances are added in, these
unitarized amplitudes usually fit the data very nicely, however, this
``solution'' has its price: the unitarizations usually ruin crossing
symmetry and analyticity, by introducing unphysical singularities into
the results, making their use for an extrapolation to lower energies
doubtful.

We have set up a framework that combines the analytic structure found
in CHPT with the constraints from unitarity:\cite{Becher:2001hv} one
starts by writing a dispersive representation for the result found in
the low energy effective theory. This representation splits the
amplitude into a polynomial part and nine functions of a {\em single}
variable, which are given by integrals over the imaginary parts of the
amplitude.  In the elastic region, unitarity then leads a set of
coupled integral equations for these functions, similar to the Roy
equations in $\pi\pi$-scattering. Replacing the imaginary parts found
in CHPT by the experimental imaginary parts in the inelastic region
and solving the equations iteratively one arrives at a representation
of the amplitude that fulfills both the constraints from unitarity and
analyticity. In addition to the imaginary parts, this system of
equations also needs four subtraction constants as an input. One of
them can be expressed as an integral over the total cross section,
while the other three need to be pinned down from the experimental
information at low energies. The results from the study of pionic
hydrogen, to be discussed below, should subject these constants to
stringent bounds.

\subsection{Isospin violation, pionic hydrogen}
To study strong isospin breaking, one needs to disentangle it from
electromagnetic isospin violation. Since both are of similar
magnitude, they need to be treated simultaneously, making it necessary
to incorporate the photon field as an additional degree of freedom
into the low energy effective Lagrangian. In the baryon sector, the
corresponding Lagrangian has been worked out to third
order\cite{Muller:1999ww} in a simultaneous expansion in $m_q\sim q^2
\sim e^2$ and the result for the pion-nucleon scattering amplitude has
been worked out to the same order.\cite{isospin}

An important application of the low energy effective theory of QCD+QED
is the extraction of the hadronic scattering length from the
measurements of the strong interaction width and level shifts of
hadronic atoms. The goal of the experiments with pionic hydrogen (the
bound state of a $\pi^-$ with a proton) at PSI \cite{PSI} is to
measure these quantities at the level of one per cent. In order to
extract the pure QCD scattering lengths from the measurements, one
needs to remove isospin breaking effects with high precision. The
framework for the calculation has been set up and by now, the
calculation of the strong energy shift has been carried out to
next-to-leading order in isospin breaking.\cite{Gasser:2002am} The
results differ significantly from earlier potential model calculations
which fail to consistently incorporate all of the interactions present
even at the leading order. At present, the main uncertainty in the
result of the effective theory is the value of the low energy constant
$f_1$, whose value is as yet unknown.

\section{Conclusions}

We have a good understanding of how chiral symmetry manifests itself
in the baryon sector. Chiral symmetry breaking effects, on the other
hand, are {\em small} and their determination from measurements is
nontrivial. The reason being that, in many cases, we cannot
directly confront the low energy theorems of the symmetry with the
experimental data taken at higher energies. In this situation, the
precise extrapolation of the data to lower energies becomes a central
issue. While the representations obtained in CHPT are not suitable for
this purpose, their analytic structure can be implemented into
a dispersive analysis.

\section*{Acknowledgments}
I would like to thank the organizers for this stimulating and pleasant
conference. This work has been sponsored by the Department of Energy under grant DE-AC03-76SF00515.

\end{document}